# Observation of a valence transition in (Pr,Ca)CoO$_3$ cobaltites: charge migration at the metal-insulator transition

José Luis García-Muñoz[1], Carlos Frontera[1], Aura J. Barón-González[1], Sergio Valencia[2], Javier Blasco[3], Ralf Feyerherm[2], Esther Dudzik[2], Radu Abrudan[4] and Florin Radu[2]

*1 Institut de Ciència de Materials de Barcelona, C.S.I.C., Campus universitari de Bellaterra, E-08193, Bellaterra, Barcelona, Spain*

*2 Hemholtz-Zentrum Berlin, BESSY, Albert-Einstein-Str. 15, 12489 Berlin, Germany*

*3 Instituto de Ciencias de Materiales de Aragón,  CSIC-Universidad de Zaragoza, c/ Pedro Cerbuna 12, 50009 Zaragoza, Spain*

*4 Institut für Experimentalphysik/Festkörperphysik, Ruhr-Universität Bochum, Germany*

X-ray absorption spectroscopy measurements in Pr$_{0.5}$Ca$_{0.5}$CoO$_3$ and (Pr,Y)$_{0.55}$Ca$_{0.45}$CoO$_3$ compositions reveal that the valence of praseodymium ions is stable and essentially +3 (Pr [$4f^{\,2}$]) in the metallic state, but abruptly changes when carriers localize approaching the oxidation state  +4 (Pr [$4f^{\,1}$]). This mechanism appears to be the driving force of the metal-insulator transition. The ground insulating state of Pr$_{0.5}$Ca$_{0.5}$CoO$_3$ is an homogeneous Co$^{3.5-\delta}$ state stabilized by a charge transfer from Pr to Co sites: $1/2 \text{Pr}^{3+} + \text{Co}^{3.5} \rightarrow 1/2\text{Pr}^{3+2\delta} + \text{Co}^{3.5-\delta}$ , with 2$\delta \approx$0.26 e$^-$.

Co oxides are attracting much attention in the condensed matter community as they present a rich variety of interesting phenomena [1-5]. Spin state transitions in oxides having $Co^{3+}$ ions is one of the most celebrated [6-10]. Different types of metal-insulator transitions (MIT) are being investigated in Ln-A-Co-O (Ln: lanthanide, A: alkaline earth) cobaltites having in common $CoO_6$ octahedra with a small energy difference between crystal field splitting and Hund's coupling energy [11-14].

Ln-A-Co-O oxides offer new frameworks to obtain remarkable magnetic and transport properties (for applications in the field of electronics) due to the ability of Co to adopt various oxidation states and electronic configurations. $Pr_{0.5}Ca_{0.5}CoO_3$ exhibits a MIT ($T_{MI}$=70 K) and is considered a "strongly correlated spin-crossover" system [12,14]. Recent experiments have proved the possibility of generating metallic domains in the insulating low temperature phase of $Pr_{0.5}Ca_{0.5}CoO_3$ by ultrafast photoexcitation, making this material of interest in the area of ultrafast optical switching devices [15]. The ionic description of the $t^5_{2g}(\sigma^*)^{0.5}$ metallic state involves the coexistence of intermediate spin (IS) $Co^{3+}$ (S=1, $t^5_{2g}e_g^1$) and low spin (LS) $Co^{4+}$ (S=1/2, $t^5_{2g}$) mobile species. Both ionic configurations are JT active and might require the creation of local lattice distortions in the localized state. A spin state transition from IS $Co^{3+}$ to diamagnetic LS $Co^{3+}$ and a charge-ordering of $Co^{3+}$ (S=0, $t^6_{2g}$) and $Co^{4+}$ sites was proposed as the origin of the MIT transition [12]. The first is widely accepted, but charge-ordering with two differentiated $Co^{3.5-\delta}$ and $Co^{3.5+\delta}$ sites has not been detected.

Several works have reported similar MIT in $(Pr_{1-y}Ln_y)_{1-x}Ca_xCoO_3$ (Ln: Sm, Tb, Y,..) perovskites for different Ca contents (x=0.3, 0.4, 0.5, etc) [16-18]. The existence of the MIT depends not only on $x$, but also on $y$ and the lanthanide species [18]. In these cases it was generally assumed that, besides the effects of temperature and hydrostatic pressure, both $x$ and $y$ parameters have a strong influence on the number of electrons in the $e_g$ orbitals [17]. According to magnetic and transport measurements, the transition was viewed as a sudden IS to LS transformation of $Co^{3+}$ atoms in the ionic description in some compounds, but also as a more gradual IS→LS crossover in other compositions [18].

However, in recent years there has been increasing evidence that the first-order MIT with x close to 0.5 occurs only for (Ln,Ln', Ca) cations with Ln=Pr [14,17-20], raising speculations about the role played by the Pr-O hybridization on the MIT in these cobaltites. Moreover a (Pr,Ca)-O bond contraction has been observed across $T_{MI}$ [20, 21, 22].

In this work we present x-ray absorption measurements at the Pr and Co sites revealing that a $Pr^{3+}/Pr^{4+}$ valence transition at $T_{MI}$ is responsible for charge localization, and also acts as the driving force for the stabilization of the $Co^{3+}$ LS state in the insulating phase of $Pr_{0.5}Ca_{0.5}CoO_3$ and related compositions. We provide direct experimental observation of a sudden spontaneous charge migration from Pr to Co sites exactly at the MIT. Electron migration stabilizes the trivalent state of Co in halfdoped $Pr_{0.5}Ca_{0.5}CoO_3$. Instead of disordered $Co^{3+}/Co^{4+}$ sites (ionic description) or $Co^{3.5+\delta}/Co^{3.5-\delta}$ sites (charge disproportionation), the actual ground state consists of an homogeneous $Co^{3.5-\delta}$ state ($\delta \approx 0.13 e^-$) arising from an electron transfer from praseodymium (electron migration from A to B sites of the perovskite).

Polycrystalline samples of $Pr_{0.5}Ca_{0.5}CoO_3$, $Pr_{0.55}Ca_{0.45}CoO_3$ and $Pr_{0.50}Y_{0.05}Ca_{0.45}CoO_3$ were prepared as detailed elsewhere [20]. Final sintering of the samples was performed under high oxygen pressure ($p_{O2}$=200 bars) to assure optimal oxygen content. Samples were well crystallized and free from impurities as revealed by x-ray and neutron diffraction. Neutron powder diffraction (NPD) data were collected on D2B [λ=1.594

Å], D1B [λ =2.52 Å] and D20 [λ=1.88 Å] diffractometers of ILL (Grenoble). Diffraction patterns were recorded in the temperature range between 5 K and room temperature (RT). Magnetization measurements were performed using a superconducting quantum interference device (SQUID) magnetometer, AC susceptibility, magnetotransport and thermoelectric power data were recorded using a commercial Physical Properties Measurement System (PPMS) in the temperature range 2<T<350 K. Absorption measurements were performed on beamlines MAGS and PGM-3 at the BESSY synchrotron radiation source [23]. X-ray absorption spectra (XAS) were collected by means of bulk-sensitive fluorescence yield (FY). Temperature dependence of the absorption was measured on heating between 10 K and 300 K.

Figure 1(a) shows the evolution with temperature of the unit cell volume of all three compounds as determined from Rietveld refinement of NPD data. $Pr_{0.55}Ca_{0.45}CoO_3$ does not show the characteristic volume contraction with decreasing T associated to the localization of the charges in these systems. This result is consistent with the temperature dependence of the resistivity (ρ) and of the susceptibility (χ) (see Figs. 1(b) and 1(c), respectively) which show that this compound remains metallic even at low temperature. Moreover, the metallic phase of $Pr_{0.55}Ca_{0.45}CoO_3$ displays long range FM order below $T_C$=70K. On the contrary, both $Pr_{0.5}Ca_{0.5}CoO_3$ and $Pr_{0.50}Y_{0.05}Ca_{0.45}CoO_3$ display a volume cell contraction concomitant with a first order MIT. The transition to a low temperature charge-localized state takes place in $Pr_{0.50}Y_{0.05}Ca_{0.45}CoO_3$ at a higher temperature ($T_{MI}$=120K) than in the half doped cobaltite ($T_{MI}$=70 K). Resistivity data agrees with the opening of a gap and both compounds exhibit semiconductor-like behaviour below $T_{MI}$.

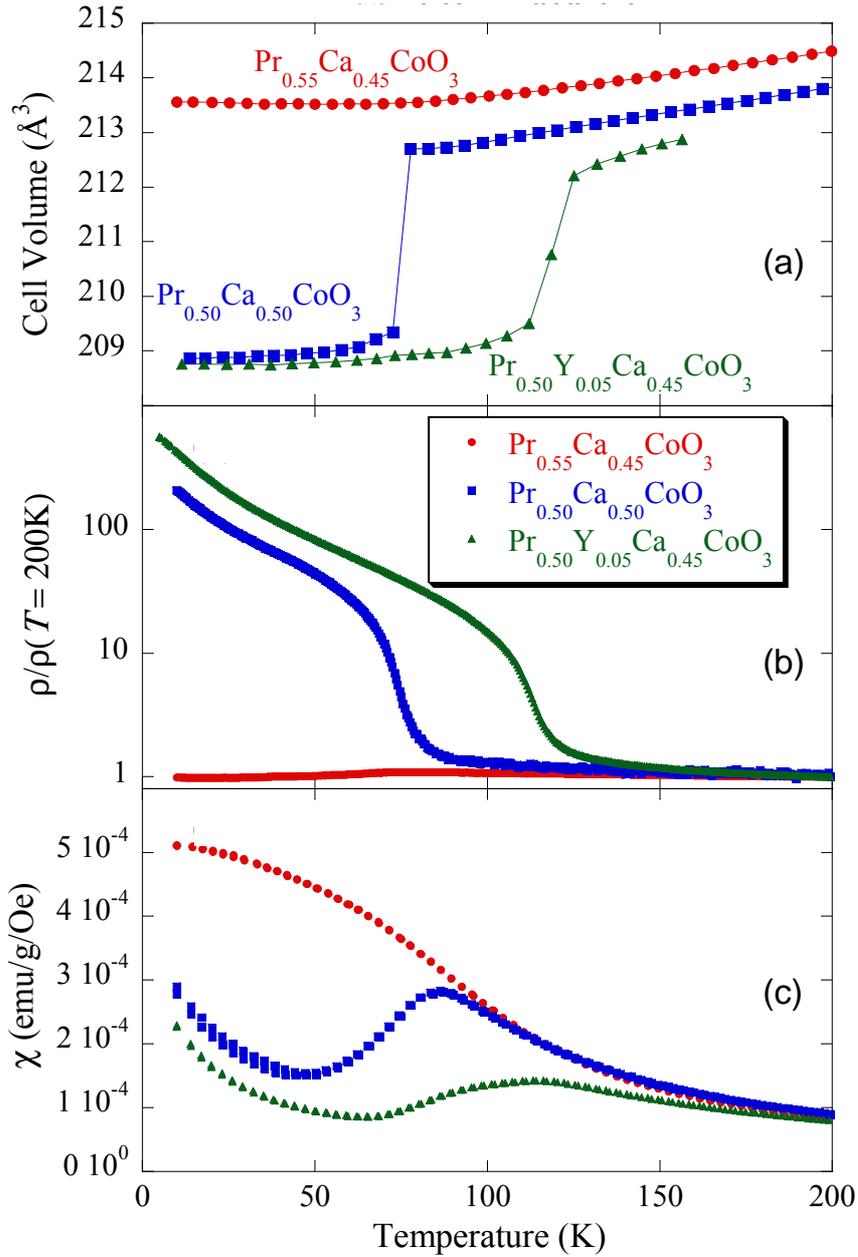

**Figure 1.** (Color online) (a) Thermal evolution of the unit cell volume for $Pr_{0.5}Ca_{0.5}CoO_3$, $Pr_{0.55}Ca_{0.45}CoO_3$ and $Pr_{0.5}Y_{0.05}Ca_{0.45}CoO_3$, obtained from neutron diffraction measurements. The volume contraction indicates the entering of the insulating phase. It is not present in $Pr_{0.55}Ca_{0.45}CoO_3$ (FM and metallic ground state). (b) Resistivity of the three perovskites plotted in logarithmic scale. (c) Magnetic susceptibility (0,1 T, zero-field cooling or cooling and heating cycle).

To investigate the role of the Pr cation on the MIT transition X-ray absorption spectra at the Pr $L_3$ edge have been acquired on MAGS (BESSY) as a function of temperature for all three samples. Absorption spectra have been obtained by means of fluorescence yield (FY-XAS). All spectra are depicted after normalization to their area on Fig. 2. The two main spectroscopic features at 5967 eV and 5978 eV (respectively A and B) originate from the Pr $2p \rightarrow 5d$ transitions: $Pr^{3+}$ ($4f^2$) sites contribute to the peak at lower energy, and the $Pr^{4+}$ sites contribution splits over the two peaks due to $4f^1$ and $4f^2\underline{L}$ states, being $\underline{L}$ a ligand hole in the O $2p$ orbital [24,25]. In $Pr_{0.5}Ca_{0.5}CoO_3$ [Fig. 2(a)] the measurements show very apparent changes in the spectra when crossing $T_{MI}$. The intensity of peak B ($Pr^{4+}$ only contribution, i.e. $4f^1$) increases at the same time as the intensity of peak A (being $Pr^{3+}$ the major contributor, i.e. $4f^2$) decreases. These results reveal a rapid increase of the Pr oxidation state at the transition. Photoabsorption spectra evidence that praseodymium changes to an intermediate valence ground state composed essentially of an atomiclike $4f^1$ state when crossing $T_{MI}$. A localized $4f^1$ configuration with a degree of covalent mixing should create in the oxygen $2p$ valence band extended states of $f$ symmetry.

The above results highlight the role of the Pr-O hybridization on the MIT. To further confirm this point Fig. 1(b) shows the Pr $L_3$ FY-XAS spectra obtained for $Pr_{0.55}Ca_{0.45}CoO_3$ at various temperatures. No relevant spectroscopic changes are seen in this case pointing to a temperature independent oxidation state of Pr. This agrees with the fact that no MIT is seen in such composition. The persistence of the metallic state in this case is related to the degree of the $GdFeO_3$-type distortion. The *Pnma* structure of $Pr_{0.55}Ca_{0.45}CoO_3$ is less distorted than $Pr_{0.5}Ca_{0.5}CoO_3$. From our NPD data one obtains the following Co-O-Co bond angles: $\theta_1$[Co-O(1)-Co] = 159.0(1)° and $\theta_2$[Co-O(2)-Co] = 157.76(1)° at RT in metallic $Pr_{0.55}Ca_{0.45}CoO_3$. This distortion should be compared with $\theta_1$=158.18(3)° and $\theta_2$=157.95(6)° in $Pr_{0.5}Ca_{0.5}CoO_3$. Although the $\theta_2$ values are approximately equal (slightly greater for the half-doped cobaltite), the apical bonding angle $\theta_1$ is by ≈1° more straight in $Pr_{0.55}Ca_{0.45}CoO_3$.

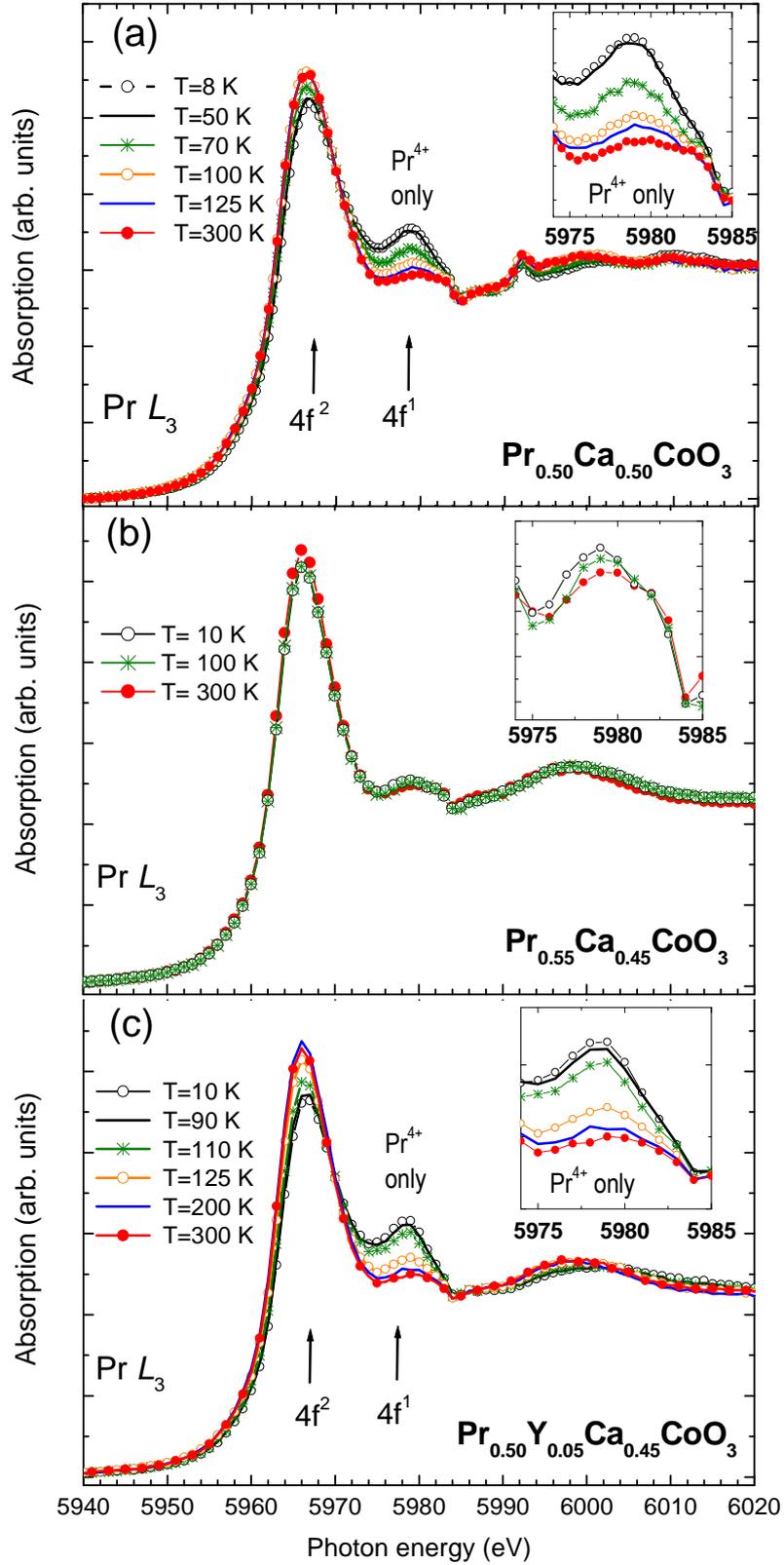

**Figure 2.** (Color online) Temperature dependence of the Pr $L_3$-edge FY-XAS of (a) $Pr_{0.5}Ca_{0.5}CoO_3$, (b) $Pr_{0.55}Ca_{0.45}CoO_3$ (metallic at all temperatures), and (c) $Pr_{0.5}Y_{0.05}Ca_{0.45}CoO_3$. Insets: enlarged $4f^1$ region.

Table I illustrates the decisive importance of the apical tilting of the octahedra ($\theta_1$) for the activation of the $4f^2/4f^1$ electronic transition. In this Table we compare the bond-lengths between A-site (Pr,Ca) and O atoms at both sides of the metal-insulator transition in $Pr_{0.5}Ca_{0.5}CoO_3$. Each praseodymium is bonded to 12 oxygens: eight basal oxygens at 8d general position (Pr-O2 bonds), and four apical oxygens at 4c positions (Pr-O1). The two strongest Pr-O bond contractions ($\approx$-2.4%) at MIT involve apical oxygens (Pr-O1). Pr-O1 bonds are parallel to the *a-c* plane because Pr and O1 atoms both occupy 4c positions (x ¼ z), at the same *y*-layer. Both atoms shift considerably their positions in the plane varying the tilting angle $\phi_1=1/2(180°-\theta_1)$ of the octahedra. The tilting angles $\theta_1$ and $\theta_2$ change by $\approx 3°$ on cooling across the transition [20-22]. The change $\Delta\theta_1$ is the main responsible of opposite spatial shifts of Pr and O1 atoms within the *a-c* plane: $u_x$=0.0079, $u_z$=-0.0023 and |s|=0.044 Å for Pr, and $u_x$= -0.0044, $u_z$= 0.0097 and |s|= 0.0567 Å for O1, where $u_i$ are the relative and |s| the total displacement.

**Table I:** Refined A-O bond-lengths in $Pr_{0.5}Ca_{0.5}CoO_3$ and relative variation across the metal-insulator transition due to $Pr^{3+}/Pr^4$ valence change. A-site is occupied by [Pr,Ca] atoms.

| $Pr_{0.5}Ca_{0.5}CoO_3$ | 10 K insulator | Relative variation 10K-100K | 100 K metal | 300 K metal |
|---|---|---|---|---|
| A-O1 (Å) | 2.286(3) | -2.36 % | 2.340(3) | 2.380(3) |
| A-O1 (Å) | 2.396(2) | -2.38 % | 2.453(2) | 2.475(2) |
| A-O2 (Å) x2 | 2.306(2) | -2.08 % | 2.354(2) | 2.344(2) |
| A-O2 (Å) x2 | 2.538(2) | -1.22 % | 2.569(2) | 2.585(2) |
| A-O2 (Å) x2 | 2.591(2) | -1.00 % | 2.617(2) | 2.630(2) |
| $<d_{A-O}>_{8-short}$ (Å) | 2.444 | | 2.484 | 2.497 |
| A-O1 (Å) | 3.009(2) | +2.29 % | 2.940(2) | 2.921(2) |
| A-O1 (Å) | 3.015(3) | +0.50 % | 3.000(3) | 3.013(3) |
| A-O2 (Å) x2 | 3.229(2) | +1.79 % | 3.171(2) | 3.139(2) |
| $<d_{A-O}>_{4-long}$ (Å) | 3.120 | | 3.070 | 3.053 |

To get further insight on the decisive importance of the tilting of the CoO$_6$ octahedra for the occurrence of the 4$f^{\,2}$→4$f^{\,1}$ transition we compare the above results with those obtained for Pr$_{0.50}$Y$_{0.05}$Ca$_{0.45}$CoO$_3$. In that case the substitution of a 5% of Y atoms, of smaller size than Pr and Ca, increases the tilting of the octahedra. From NPD we obtain $\theta_1$ = 157.56(1) ° and $\theta_2$ =157.67(1)° at RT. The basal angle $\theta_2$ is almost identical to metallic Pr$_{0.55}$Ca$_{0.45}$CoO$_3$ but the apical angle $\theta_1$ is ≈1.5° more bended. The Pr L$_3$ FY-XAS spectra for this sample (Fig. 1(c)) indeed show, as did for Pr$_{0.50}$Ca$_{0.50}$CoO$_3$, a very clear evolution where the Pr$^{4+}$ contribution increases below ≈130K. In this case the tilting angle $\phi_1$ brings the apical oxygen O1 nearer to some Pr atom at the y=1/4 layer. Moreover, $\phi_1$ tilts the basal plane CoO2$_4$ of the octahedron towards a more vertical position moving the basal oxygens O2 towards the up (y=1/4) and down (y=3/4) Pr planes.

The accuracy of the FY-XAS technique allows deriving a quantitative description of the valence changes with great precision. As in some previous XAS studies of L$_3$ edges (e.g. Yamaoka et al [25] and Richter et al. [26]) the Pr valence is estimated from the relative ratio of intensities corresponding to A and B spectral features, i.e. I$_A$/I$_B$ [26]. Both I$_A$ and I$_B$ were obtained by fitting the spectra with two Voigt functions after subtracting the step-like signal from the continuum. The temperature evolution of the Pr valence for all three compositions is plotted in Fig. 3. The absorption data at ambient temperature indicate an essentially trivalent Pr in all cases however already presenting a small hybridization with O 2$p$ states. This is consistent with electronic-structure calculations reported by Knížek *et al*, from which a certain Pr(4$f$)-O(2$p$) hybridization produces a broad band 0.5 eV below Fermi level, much broader than corresponding Nd(4$f$) band in Nd$_{0.5}$Ca$_{0.5}$CoO$_3$ [14]. Decreasing temperature, between RT and T$_{MI}$, the system keeps its metallic state and according to Fig. 3 the valence of Pr remains constant. As the temperature approaches $T_{MI}$, the Pr valence rapidly increases: from v=3.33 to 3.60 in $Pr_{0.5}Ca_{0.5}CoO_3$ ($\Delta v[Pr]$= 0.27(2)$e^-/_{Pr}$). Very similar within errors is the valence change of praseodymium in $Pr_{0.50}Y_{0.05}Ca_{0.45}CoO_3$ ($\Delta v[Pr]$= 0.26(2)$e^-/_{Pr}$). The evolution below RT corresponding to metallic $Pr_{0.55}Ca_{0.45}CoO_3$ is also shown in Fig. 3. This compound does not show, within errors, appreciable changes in the electronic configuration of Pr. The inset of Fig. 3 displays a comparison of the valence at Pr sites with the Seebeck coefficient ($\alpha$) as a function of temperature for $Pr_{0.5}Ca_{0.5}CoO_3$. Changes in the oxidation state of praseodymium are perfectly simultaneous to changes in the Seebeck coefficient.

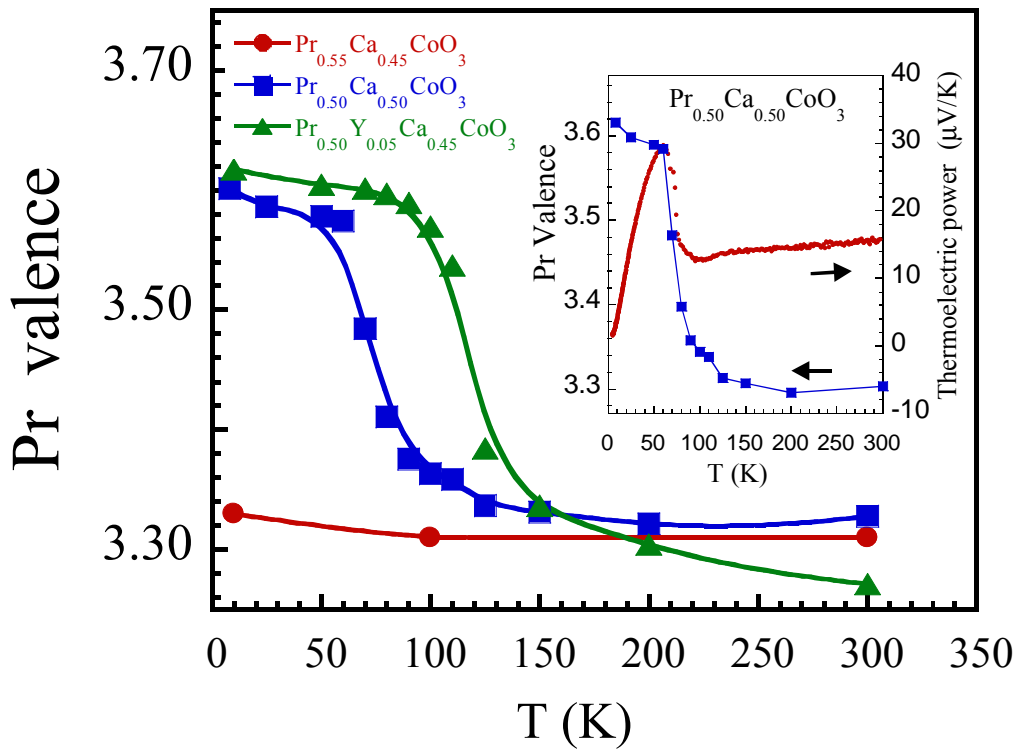

**Figure 3.** (Color online) Comparative thermal evolution of the valence of praseodymium across the MIT in $Pr_{0.5}Ca_{0.5}CoO_3$ and $Pr_{0.5}Y_{0.05}Ca_{0.45}CoO_3$, and in metallic $Pr_{0.55}Ca_{0.45}CoO_3$. The solid lines are guides to the eye. Inset: Comparison of the thermal dependences of the Pr valence and the Seebeck coefficient in $Pr_{0.5}Ca_{0.5}CoO_3$.

So, XAS data reveals that the strong Pr-O bond contraction detected at $T_{MI}$ by neutron diffraction is due to a significant overlap between the Pr 4$f$ wave functions and selected oxygens that abruptly shifts Pr valence towards the tetravalent 4$f^1$ state [27]. As the electrical neutrality requires, Co $L_3$ XAS spectra of $Pr_{0.5}Ca_{0.5}CoO_3$ confirm that the electronic migration from Pr is compensated by changes in the Co valence. Co $L_3$ XAS spectra recorded by using the Alice diffractometer chamber [28] on PGM-3 are shown in Fig. 4 as a function of temperature. Important modifications show up at this edge coinciding with the transition. As a guide for the valence states of Co ions, the changes in Fig. 4 are compared to the reference compounds $SrCoO_3$ ($Co^{4+}$) and $AgCoO_2$ ($Co^{3+}$) (from Ref. 29). The spectra of these formally tetravalent and trivalent systems at RT mainly differ in the intensity of the shoulder at 2eV below the main peak. A clear loss of intensity in the shoulder at ~779 eV (2 eV below the maximum) has been detected at MIT in $Pr_{0.5}Ca_{0.5}CoO_3$ as expected from a charge transfer from Pr to Co sites. Outside the temperature window of the transition the spectra shown in Fig. 4 hardly change.

Neutron diffraction data and XAS spectra are thus consistent and indicate that the average valence of cobalt should decrease approaching the trivalent state. Moreover, the $Co^{3+}$ IS state and the HS are not favored in the low temperature phase because the smaller cell volume favors enhanced crystal-field splitting between $e_g$ and $t_{2g}$ orbitals [29]. As a result, the detected transfer of charges in $Pr_{0.5}Ca_{0.5}CoO_3$ should correspond to:

½ [Pr]4$f^2$ + ½ [Co]$t^5_{2g}e^1_g$ + ½ [Co]$t^5_{2g}$ → ½ [Pr]4$f^1$ + [Co]$t^6_{2g}$   per formula unit (f. u.).

Namely, there is a stabilization of the trivalent LS state of Co in the localized phase induced by a destabilization of the trivalent valence state of praseodymium. According to Fig. 3, electron migration is only partial: 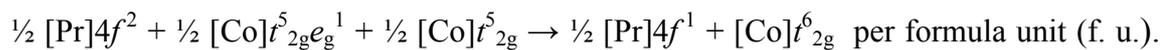 1/2$Pr^{3+}$ + $Co^{3.5}$ → 1/2$Pr^{3+2\delta}$ + $Co^{3.5-\delta}$ , with 2$\delta$=Δv[Pr]= ≈0.26 e$^-$. The drop in the susceptibility below $T_{MI}$ is mainly a signature of the stabilization of the $t^6_{2g}$ state at $Co^{3+}$ sites (LS), but it is also due to a reduction of the moment at Pr sites.

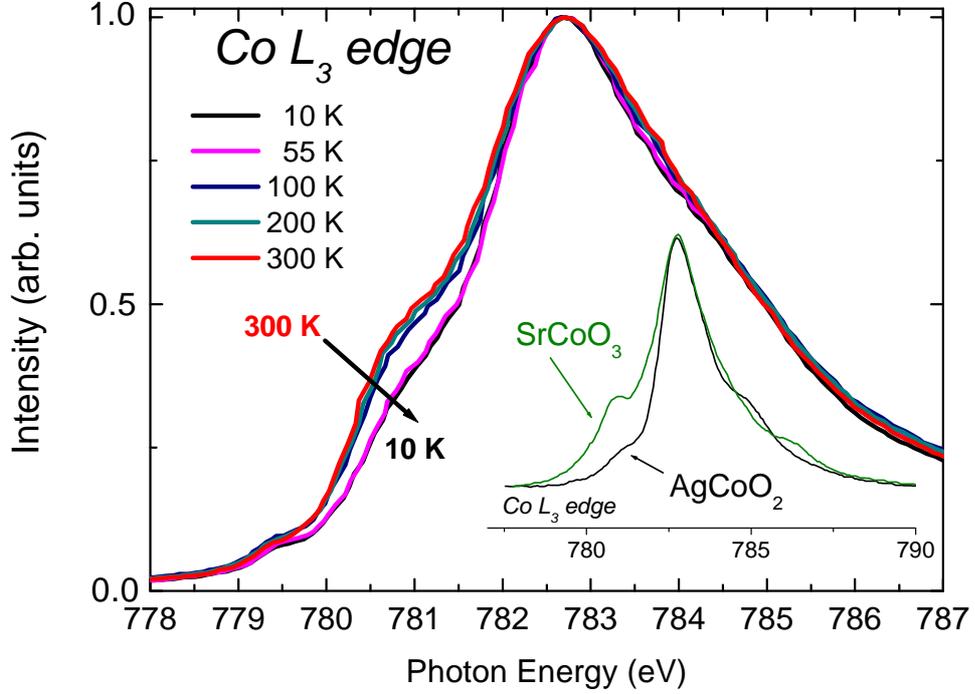

**Figure 4.** (Color online) Co $L_3$ edge spectra of $Pr_{0.5}Ca_{0.5}CoO_3$ across the transition. The RT spectra of $SrCoO_3$ and $AgCoO_2$ (ref. 30) are shown for comparison.

The present experimental results nicely agree with the electronic structure calculations by Knížek et al in Ref. 14. They predicted a reduction of $t^5_{2g}(\sigma^*)^{0.5}$ states around the Fermi level, and their substitution at the critical temperature by a narrow Pr 4$f$ band [14]. They anticipated that the gap could be associated to a charge transfer between Pr and Co sites. Absorption data across the Pr $L_3$ edge are consistent with the energy of Pr 4$f$ bands being relatively near the Fermi level in the metallic state, and a part of them becoming unoccupied and shifted above $E_F$ below $T_{MI}$. Simultaneously, concomitant with the gap opening, some initially empty Co $t_{2g}$ symmetry states move below Fermi level being occupied by electrons originally in states of $f$ symmetry. The result is a charge transfer from Pr-4$f$ to Co-3$d$-$t_{2g}$ states at $T_{MI}$ on cooling.

In summary, X-ray absorption spectroscopy measurements in $Pr_{0.5}Ca_{0.5}CoO_3$ and $(Pr,Y)_{0.55}Ca_{0.45}CoO_3$ compositions reveal that the valence of praseodymium ions is stable and essentially +3 (Pr $[4f^2]$) in the metallic state, but abruptly increases approaching the oxidation state +4 (Pr $[4f^1]$) at the MIT transition. This mechanism appears to be the driving force of the metal-insulator transition. Moreover, the tilting $\phi_1$ of $CoO_6$ octahedra is found to play a prominent role for the occurrence of the valence change and the concomitant electron localization.

Electrons leaving Pr sites are used to stabilize the trivalent low-spin state of Co. The ground state at low temperature is not realized through segregation of low- or intermediate-spin $Co^{3+}$ and low-spin $Co^{4+}$ ionic species. Instead, the $t^5_{2g}(\sigma^*)^{0.5}$ metallic state is substituted by an homogeneous $Co^{3.5-\delta}$ state stabilized by electron transfer from praseodymium: $1/2Pr^{3+} + Co^{3.5} \rightarrow 1/2Pr^{3+2\delta} + Co^{3.5-\delta}$, with $2\delta \approx 0.26 e^-$ in $Pr_{0.5}Ca_{0.5}CoO_3$.

**Acknowledgements**

Financial support from MICINN (Spanish government) under projects MAT2006-11080-C02-02 and No. MAT2009-09308, and NANOSELECT under Project No. CSD2007-00041 is thanked. We thank ILL (and the CRG-D1B), and BESSY for the provision of beamtime. The authors gratefully acknowledge C. Ritter for his assistance with neutron data collection.